\documentclass[11pt]{article}
\usepackage{fullpage}
\usepackage{cite}
\usepackage{graphicx}
\usepackage{amsmath}
\usepackage{amssymb}
\title{}
\date{}

\def\beq{\begin{equation}}
\def\eeq{\end{equation}}
\allowdisplaybreaks
\begin{document}
\bibliographystyle{utphys}
\newcommand{\msbar}{\ensuremath{\overline{\text{MS}}}}
\newcommand{\DIS}{\ensuremath{\text{DIS}}}
\newcommand{\abar}{\ensuremath{\bar{\alpha}_S}}
\newcommand{\bb}{\ensuremath{\bar{\beta}_0}}
\newcommand{\rc}{\ensuremath{r_{\text{cut}}}}
\newcommand{\Nd}{\ensuremath{N_{\text{d.o.f.}}}}

\newcommand{\cdf}{{D}}
\newcommand{\pa}{{\partial}}
\newcommand{\pd}{{\partial}}

\setlength{\parindent}{0pt}

\titlepage
\begin{flushright}
QMUL-PH-17-16
\end{flushright}

\vspace*{0.5cm}

\begin{center}
{\bf \Large The Kerr-Schild double copy in curved spacetime}

\vspace*{1cm}
\textsc{Nadia Bahjat-Abbas $^a$ \footnote{n.bahjat-abbas@qmul.ac.uk},
Andr\'{e}s Luna $^b$ \footnote{a.luna-godoy.1@research.gla.ac.uk} 
and Chris D. White $^a$ \footnote{christopher.white@qmul.ac.uk}} \\

\vspace*{0.5cm} $^a$ Centre for Research in String Theory, School of
Physics and Astronomy, \\
Queen Mary University of London, 327 Mile End
Road, London E1 4NS, UK\\

\vspace*{0.5cm} $^b$ School of Physics and Astronomy\\ University of
Glasgow, Glasgow G12 8QQ, Scotland, UK

\end{center}

\vspace*{0.5cm}

\begin{abstract}
The double copy is a much-studied relationship between scattering
amplitudes in gauge and gravity theories, that has subsequently been
extended to classical field solutions. In nearly all previous
examples, the graviton field is defined around Minkowski
space. Recently, it has been suggested that one may set up a double
copy for gravitons defined around a non-trivial background. We
investigate this idea from the point of view of the classical double
copy. First, we use Kerr-Schild spacetimes to construct graviton
solutions in curved space, as double copies of gauge fields on
non-zero gauge backgrounds. Next, we find that we can reinterpret such
cases in terms of a graviton on a non-Minkowski background, whose
single copy is a gauge field in the same background spacetime. The
latter type of double copy persists even when the background is not of
Kerr-Schild form, and we provide examples involving conformally flat
metrics. Our results will be useful in extending the remit of the
double copy, including to possible cosmological applications.
\end{abstract}

\vspace*{0.5cm}

\section{Introduction}
\label{sec:intro}

The structure of gauge and gravity theories, and the relationships
between them, continue to be the subjects of active research. An
important example of such a relationship is the {\it double
  copy}~\cite{Bern:2008qj,Bern:2010ue,Bern:2010yg}, which states that
perturbative scattering amplitudes in gravity theories both with and
without supersymmetry can be obtained from their counterparts in a
non-abelian gauge theory by exchanging the couplings, and also
replacing colour information with kinematic information in a
prescribed way. This relies on a certain colour-kinematics interplay
-- {\it BCJ duality} -- being possible in the gauge
theory~\cite{Bern:2008qj}. Both BCJ duality and the double copy are
proven at
tree-level~\cite{BjerrumBohr:2009rd,Stieberger:2009hq,Bern:2010yg,BjerrumBohr:2010zs,Feng:2010my,Tye:2010dd,Mafra:2011kj,Monteiro:2011pc,BjerrumBohr:2012mg},
where the latter is equivalent to the known KLT
relations~\cite{Kawai:1985xq} between gauge and gravity amplitudes,
that arise from string theory. At loop level these properties remain
conjectural, although highly non-trivial evidence exists at multiloop
level in a variety of
theories~\cite{Bern:2010ue,Bern:1998ug,Green:1982sw,Bern:1997nh,Carrasco:2011mn,Carrasco:2012ca,Mafra:2012kh,Boels:2013bi,Bjerrum-Bohr:2013iza,Bern:2013yya,Bern:2013qca,Nohle:2013bfa,
  Bern:2013uka,Naculich:2013xa,Du:2014uua,Mafra:2014gja,Bern:2014sna,
  Mafra:2015mja,He:2015wgf,Bern:2015ooa,
  Mogull:2015adi,Chiodaroli:2015rdg,Bern:2017ucb}. All-order evidence
is possible in certain special kinematic
limits~\cite{Oxburgh:2012zr,White:2011yy,Melville:2013qca,Luna:2016idw,Saotome:2012vy,Vera:2012ds,Johansson:2013nsa,Johansson:2013aca},
and other related studies can be found
in~\cite{Monteiro:2013rya,Tolotti:2013caa,Fu:2013qna,Du:2013sha,Fu:2012uy,Naculich:2014naa,Naculich:2014rta,Chiodaroli:2014xia,Carrasco:2013ypa,Litsey:2013jfa,Nagy:2014jza,Anastasiou:2015vba,Johansson:2015oia,Lee:2015upy,Barnett:2014era,Chiodaroli:2016jqw,Cardoso:2016ngt,Mizera:2016jhj,Bern:2017yxu,Chiodaroli:2017ngp,Johansson:2017bfl,Johansson:2017srf,Cheung:2016say,Cheung:2017kzx,Cheung:2017ems}. \\

It remains unclear whether or not the double copy is an accident of
perturbative scattering amplitudes, or represents a much deeper
relationship between gauge, gravity and related theories. A number of
recent studies have therefore looked at whether or not other
quantities can be matched up and, if so, whether the relevant relationship
is related to the double copy for
amplitudes. Reference~\cite{Monteiro:2014cda} considered a family of
exact classical solutions in gravity, {\it stationary Kerr-Schild
  metrics}, and found that these could indeed be single-copied to
Yang-Mills theory. Well-known gravitational objects such as the
Schwarzschild and Kerr black holes emerge as special cases. The
Kerr-Schild framework can also be generalised to include
(time-dependent) plane waves, and known properties of amplitudes in
the self-dual sector of Yang-Mills theory and
gravity~\cite{Monteiro:2011pc}. The case of an arbitrarily
accelerating particle was considered in ref.~\cite{Luna:2016due}. This
also has a Kerr-Schild form, which has the effect of forcing the
radiation to appear as a source term on the right-hand side of the
Yang-Mills and Einstein equations. This source term could be related
to known amplitudes for the Bremsstrahlung of photon and graviton
radiation, thus more tightly establishing the link between the
classical and amplitude double copies. \\

Reference~\cite{Luna:2015paa} went beyond the simple Kerr-Schild form
used in ref.~\cite{Monteiro:2011pc}, in considering the Taub-NUT
solution. This has a double Kerr-Schild form, yet nevertheless can be
single copied to a gauge theory dyon, whose magnetic monopole charge
maps to the NUT charge in the gravity theory. Furthermore,
ref.~\cite{Luna:2015paa} already hinted at the possibility of
constructing a double copy around a non-trivial background metric
(namely the de Sitter metric), and we will return to this in what
follows. Further work relating to the classical double copy has
investigated whether or not the source terms in the field equations
are physically meaningful~\cite{Ridgway:2015fdl}, and whether the copy
can be extended to solutions involving inverse powers of the
coupling~\cite{White:2016jzc,DeSmet:2017rve}. An alternative body of
work has focused on constructing gravity fields from convolutions of
gauge fields, in a variety of
theories~\cite{Anastasiou:2014qba,Borsten:2015pla,Anastasiou:2016csv,Anastasiou:2017nsz,Cardoso:2016ngt,Borsten:2017jpt}. The
double copy has also been applied to classical solutions generated
order-by-order in perturbation
theory~\cite{Goldberger:2016iau,Goldberger:2017frp,Luna:2016hge}.\\

Recently, ref.~\cite{Adamo:2017nia} considered generalising the double
copy for amplitudes to include a non-trivial background metric in the
gravity theory. Let us define the graviton $h_{\mu\nu}$ according to
\begin{equation}
g_{\mu\nu}=\bar{g}_{\mu\nu}+\kappa h_{\mu\nu},
\label{hdef}
\end{equation}
where $g_{\mu\nu}$ is the full metric, and $\kappa=\sqrt{32\pi G_N}$,
with $G_N$ Newton's constant. In the standard BCJ double copy for
amplitudes~\cite{Bern:2008qj,Bern:2010ue,Bern:2010yg}, one identifies
the background metric $\bar{g}_{\mu\nu}$ with the Minkowski metric
$\eta_{\mu\nu}$, so that zero graviton corresponds to the complete
absence of gravity. One may instead consider a different choice, and
indeed there may be good practical reasons for doing so: in a variety
of astrophysical and / or cosmological applications, one must analyse
perturbations around a non-trivial background
metric. Reference~\cite{Adamo:2017nia} considered the particular case
of so-called {\it sandwich plane waves}, namely plane wave solutions
whose deviation from Minkowski space has a finite extent in space and
time~\footnote{More precisely, such waves are confined to a finite
  region of the lightcone coordinate $u=z-t$, for a wave travelling in
  the $+z$ direction.}. One may consider such waves in either gauge
theory or gravity, and the authors demonstrate explicitly that a
three-point amplitude for a graviton defined as the deviation from a
gravitational sandwich wave can be obtained as the double copy of a
gauge theory three-point function, where the gauge field is defined
around a gauge theory sandwich wave. They further note that this
procedure is obtainable from ambitwistor string
theory~\cite{Adamo:2017sze} (see
also~\cite{Mason:2013sva,Adamo:2014wea}), which would in principle
provide a general framework for formulating a similar procedure for
different types of background. \\

Given the previously observed links between the Minkowski space
amplitude double copy of
refs.~\cite{Bern:2008qj,Bern:2010ue,Bern:2010yg} and the classical
double copy of
refs.~\cite{Monteiro:2014cda,Luna:2015paa,Luna:2016due}, the results
of ref.~\cite{Adamo:2017nia} suggest that some analogue of the curved
space amplitude double copy should also be possible for classical
solutions. The aim of this paper is to study this issue, and we will
present a number of examples. Firstly, we will construct Kerr-Schild
solutions on a curved background by trivially rewriting single
Kerr-Schild solutions. We will be able to interpret such solutions as
double copies of gauge fields with non-trivial backgrounds, and we
will call this relationship a {\it type A} curved space double
copy. We will also find an alternative interpretation, namely that one
may regard the graviton as being the double copy of a gauge field
living on a non-dynamical curved spacetime background, which we will
refer to as {\it type B}.
\begin{figure}
\begin{center}
\scalebox{1.0}{\includegraphics{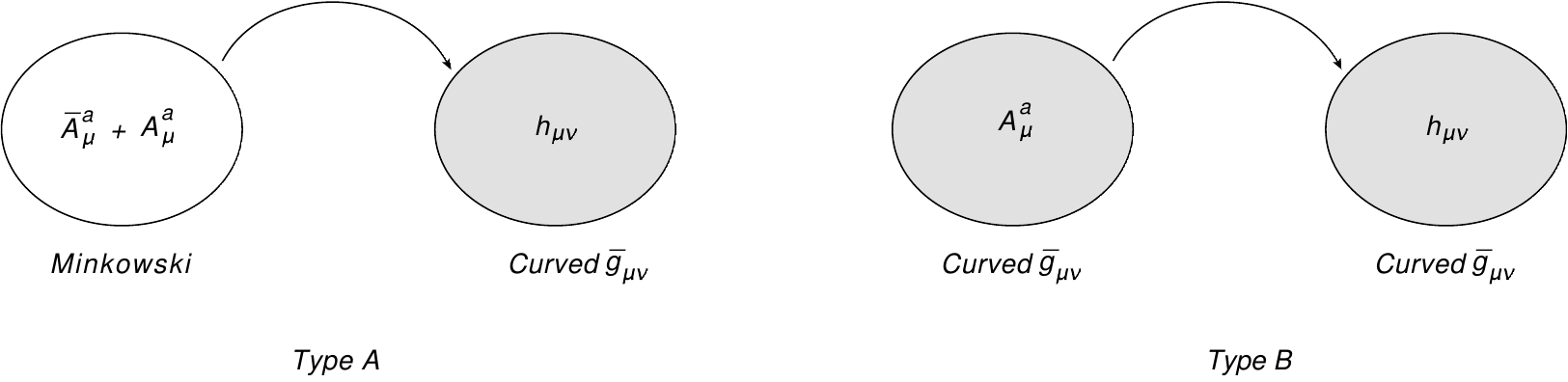}}
\caption{Two possible interpretations of a double copy in curved
  space: in {\it type A}, a gauge field has a non-trivial background
  field $\bar{A}_\mu^a$ in Minkowski space, and copies to a graviton
  defined on a curved background $\bar{g}_{\mu\nu}$, where
  $\bar{g}_{\mu\nu}$ and $\bar{A}_\mu^a$ are themselves related by a
  double copy relationship. In {\it type B}, a gauge field on a
  non-dynamical curved background $\bar{g}_{\mu\nu}$ double copies to
  a graviton defined around the same background.}
\label{fig:curvedmap}
\end{center}
\end{figure}
The difference between these two double copies is shown schematically
in figure~\ref{fig:curvedmap}, and the second of these is perhaps at
odds with what one normally means by the double copy, which relates
entire gravity solutions to gauge theory counterparts in flat
space. It is then presumably the case that the type B map is not
fully general, but exists only in special cases. That does not
however, reduce its usefulness, where it applies.\\

After examining simple Kerr-Schild examples, we will generalise our
findings to multiple Kerr-Schild solutions, including a reexamination
of the Taub-NUT spacetime considered in
ref.~\cite{Luna:2015paa}. Finally, we will show a family of
non-trivial examples of the type B double copy map, in which the
background spacetime is conformally flat, without a Kerr-Schild
form. This illustrates that this second type of double copy map may be
more applicable than na\"{i}vely thought, and can also provide a
double copy in cases in which it is not known how to construct a
double copy of type A.\\

The structure of the paper is as follows. In section~\ref{sec:review},
we briefly review relevant details regarding the classical double copy
that will be needed in what follows. In section~\ref{sec:KS}, we study
Kerr-Schild solutions from the viewpoint of a curved space double
copy. In section~\ref{sec:Einstein}, we consider the example of
Kerr-Schild solutions built upon conformally flat background
metrics. Finally, in section~\ref{sec:discuss}, we discuss our results
and conclude. Technical details are contained in the appendix.

\section{The classical double copy}
\label{sec:review}

Here, we briefly review the double copy for classical solutions of
refs.~\cite{Monteiro:2014cda,Luna:2015paa,Luna:2016due}. Given that
this will be the focus of our paper, we will not discuss in detail the
corresponding story for amplitudes (see
e.g.~\cite{Carrasco:2015iwa,Cheung:2017pzi,White:2017mwc} for
pedagogical reviews). Our starting point is to consider {\it
  Kerr-Schild} metrics, in which the graviton, defined in
eq.~(\ref{hdef}), is given by
\begin{equation}
h_{\mu\nu}=\frac{\kappa}{2}\phi k_\mu k_\nu.
\label{hKS}
\end{equation}
Here $\phi(x^\mu)$ is a scalar function, and $k_\mu$ is a 4-vector
that is null and geodetic both with respect the background and the
whole metric. That is
\begin{equation}
g^{\mu\nu} k_\mu k_\nu = \bar{g}^{\mu\nu}k_\mu k_\nu=0,\quad
k^\mu \cdf_\mu k_\nu=0,
\label{Geodetic}
\end{equation}
where $\cdf_\mu$ is the covariant derivative associated with the
background metric, i.e. $\cdf_\rho(\bar{g}_{\mu\nu})=0$. The null
property implies that the index of the Kerr-Schild vector $k_\mu$
can be raised with either $\bar{g}^{\mu\nu}$ or $g^{\mu\nu}$, where
the inverse metric takes the simple form
\begin{eqnarray}
g^{\mu\nu}=\bar{g}^{\mu\nu}-\frac{\kappa^2}{2} k^\mu k^\nu \phi.
\label{Inverse}
\end{eqnarray}
Upon substituting the ansatz of eq.~(\ref{hdef}) into the Einstein
equations, one finds a linear form for the Ricci tensor with the
particular index placement shown below:
\begin{eqnarray}
R^\mu_{\ \nu}=\bar{R}^\mu_{\ \nu}-\kappa h^\mu_{\ \rho} \bar{R}^\rho_{\ \nu}+\frac{\kappa}{2}\cdf_\rho\left(\cdf_\nu h^{\mu\rho}+\cdf^\mu h^\rho_{\ \nu}-\cdf^\rho h^\mu_{\ \nu}\right),
\label{MixedRicci}
\end{eqnarray}
where $\bar{R}_{\mu\nu}$ is the Ricci tensor associated with
$\bar{g}_{\mu\nu}$. Thus, if one finds a field $\phi$ and Kerr-Schild
vector $k_\mu$ such that the Einstein equations (vacuum or otherwise)
are solved, this constitutes an exact solution i.e. the graviton
receives no higher order corrections. \\

Equation~(\ref{MixedRicci}) simplifies in the case that the background
space is taken to be Minkowski
\begin{equation}
R^\mu_{\ \nu}\xrightarrow{\bar{g}_{\mu\nu}\,\rightarrow\,\eta_{\mu\nu}}
\frac{\kappa}{2}\partial_{\rho}\left(
\partial_\nu h^{\mu\rho}+\partial^\mu h^\rho_{\ \nu}
-\partial^\rho h^\mu_{\ \nu}\right). 
\label{MixedRicci2}
\end{equation}
For a given field $\phi$ and Kerr-Schild vector $k^\mu$, one may
define a (non-abelian) gauge field according to
\begin{eqnarray}
A_\mu^a=c^a\phi k_\mu,
\label{SingleCopy}
\end{eqnarray}
where $c^a$ is an arbitrary constant colour vector. For all stationary
Kerr-Schild metrics (i.e. those not depending explicitly on time),
this gauge field solves the linearised Yang-Mills equations:
\begin{eqnarray}
\pd^\mu F^a_{\mu\nu}=0,
\label{FlatMaxwell}
\end{eqnarray}
where $F^a_{\mu\nu}$ is the abelian-like field strength
tensor~\footnote{The non-linear term that is usually present in the
  non-abelian field strength vanishes upon substituting the ansatz of
  eq.~(\ref{SingleCopy}).}
\begin{equation}
F^a_{\mu\nu}=\partial_\mu\,A^a_\nu-\partial_\nu\,A^a_\mu.
\label{FS}
\end{equation}
To see this, note that eqs.~(\ref{hKS}, \ref{MixedRicci2}) imply
\begin{equation}
R^0_{\ i}\xrightarrow{\bar{g}_{\mu\nu}\,\rightarrow\,\eta_{\mu\nu}}
\frac{\kappa}{2}\partial_{\rho}\left[
\partial_i (\phi \,k^0\, k^\rho)-\partial^\rho(\phi\, k^0\, k_i)\right]
\label{R0nu}
\end{equation}
where we use latin indices to denote spatial components. Without loss
of generality, one may choose a coordinate system such that $k^0=1$,
such that the result of eq.~(\ref{FlatMaxwell}) follows. As in
gravity, the linearisation of the field equations is exact, so that
the gauge field receives no higher order corrections.\\

The gauge field of eq.~(\ref{SingleCopy}) is referred to as the {\it
  single copy} of its corresponding Kerr-Schild graviton $h_{\mu\nu}$,
by analogy with the BCJ double copy for
amplitudes~\cite{Bern:2008qj,Bern:2010ue,Bern:2010yg}. In fact, the
two double copies are related to each other. One way to see this is
take the {\it zeroth copy} of eq.~(\ref{SingleCopy}), which involves
replacing the remaining copy of the vector $k_\mu$ with a second
colour vector:
\begin{equation}
\Phi^{aa'}=c^a\,\tilde{c}^{a'}\,\phi.
\label{ZerothCopy}
\end{equation}
This is a solution of a {\it biadjoint scalar} field theory, whose
equation is
\begin{equation}
\partial^2\Phi^{aa'}-yf^{abc}\tilde{f}^{a'b'c'}\Phi^{bb'}\Phi^{cc'}=0,
\label{biadjoint}
\end{equation}
where $f^{abc}$ and $\tilde{f}^{a'b'c'}$ are structure constants
associated with two (potentially different) Lie groups. Indeed,
eq.~(\ref{ZerothCopy}) is such that the nonlinear term in
eq.~(\ref{biadjoint}) vanishes, leaving the simpler equation
\begin{equation}
\partial^2\Phi^{aa'}=0.
\label{biadjoint2}
\end{equation}
When sources are present on the right-hand side, we can then interpret
$\phi$ as a scalar propagator, integrated over the source
distribution. The fact that one does not modify the function $\phi$
upon taking the single or zeroth copies of eqs.~(\ref{hKS},
\ref{SingleCopy}) is similar to the fact that denominators of
amplitudes (themselves interpretable as scalar propagators) remain the
same in biadjoint scalar, gauge and gravity theories. As mentioned
above, stronger evidence for the connection between the classical and
amplitude double copies comes from considering the accelerated
particle, where the Kerr-Schild description recovers known amplitudes
for the emission of soft photons and gravitons~\cite{Luna:2016due}.\\

It is not known how to fully generalise the classical double copy to
exact solutions which are not of Kerr-Schild form. One way to make
progress is to construct solutions order-by-order in a perturbation
expansion in $\kappa$, as explored in
refs.~\cite{Goldberger:2016iau,Goldberger:2017frp,Luna:2016hge}. One
may also explore known generalisations of the Kerr-Schild ansatz. An
example is the {\it double Kerr-Schild} ansatz, in which the graviton
has the form
\begin{equation}
h_{\mu\nu}=\frac{\kappa}{2}\Big[
\phi_1k_\mu k_\nu+\phi_2 l_\mu l_\nu.
\Big]
\label{doubleKS}
\end{equation}
There are now two scalar fields $\phi_i$, and two separate Kerr-Schild
vectors $k^\mu$ and $l^\mu$, each of which satisfies the null and
geodetic requirements of eq.~(\ref{Geodetic}), as well as the mutual
orthogonality condition
\begin{equation}
g^{\mu\nu} k_\mu l_\nu = \bar{g}^{\mu\nu}k_\mu l_\nu=0.
\label{orthogonality}
\end{equation}
Unlike in the single Kerr-Schild case, this ansatz is not guaranteed
to linearise the Einstein equations: one finds a correction term to
eq.~(\ref{MixedRicci}), whose explicit form (in the present notation)
may be found in ref.~\cite{Luna:2015paa}. A special case where
linearity indeed occurs is the Taub-NUT solution~\cite{Taub,NUT},
whose Kerr-Schild form was first presented in
ref.~\cite{Chong:2004hw}. The two terms in eq.~(\ref{doubleKS})
contain, respectively, a Schwarzschild-like point mass $M$ at the
origin, and a {\it NUT charge} $N$, where the latter gives rise to a
rotational character in the gravitational field at spatial
infinity. Reference~\cite{Luna:2015paa} single copied this solution by
analogy with eq.~(\ref{SingleCopy}):
\begin{equation}
A^a_\mu=c^a\Big[
\phi_1\, k_\mu+\phi_2\, l_\mu
\Big].
\label{SingleCopy2}
\end{equation}
Note that the single copy is taken term-by-term, analogous to the BCJ
double copy for amplitudes. The gauge theory solution was found to be
a dyon. The electric charge in the gauge theory maps to the
Schwarzschild mass, as it must do for consistency with the pure
Schwarzschild case. The magnetic monopole charge of the gauge theory
solution maps to the NUT charge, thus making precise the statement
that the Taub-NUT solution can be thought of as magnetic-monopole-like. \\

Of particular interest for the present study is the fact that
ref.~\cite{Luna:2015paa} considered the Taub-NUT solution on a de
Sitter background, as well as Minkowski space. The corresponding gauge
field was then found to satisfy the {\it curved space} Maxwell
equations
\begin{equation}
D^\mu F^a_{\mu\nu}=0,
\label{Maxwell_curved}
\end{equation}
where $D^\mu$ is the covariant derivative associated with de Sitter
space, and
\begin{equation}
F^a_{\mu\nu}\equiv D_\mu A^a_\nu-D_\nu A^a_{\mu}
=\partial_\mu A^a_\nu-\partial_\nu A^a_{\mu}
\label{FScurved}
\end{equation}
is the curved space field strength tensor~\footnote{The second
  equality in eq.~(\ref{FScurved}) follows from the fact that terms
  involving the Christoffel symbol vanish upon forming the
  antisymmetric combination of covariant derivatives.}. This is
already an example of the type B double copy illustrated in
figure~\ref{fig:curvedmap}, in which a graviton defined around a
non-Minkowski background is identified with a gauge field living on
the same (non-dynamical) background. Given the fact that this map is
not what one ordinarily associates with the double copy, it is not
clear what, if at all, the zeroth copy of the gauge field corresponds
to. For the Taub-NUT example of ref.~\cite{Luna:2015paa}, the
biadjoint field
\begin{equation}
\Phi^{aa'}=c^a\,\tilde{c}^{a'}\Big(\phi_1+\phi_2\Big)
\label{Taub-NUT-zeroth}
\end{equation}
was found to satisfy the equation
\begin{equation}
D^2 \Phi^{aa'}\equiv D^\mu D_\mu \Phi^{aa'}=-2\lambda\Phi^{aa'},
\label{TNPhieq}
\end{equation}
where $\lambda$ is the cosmological constant. It was speculated that
this was a solution of the biadjoint theory conformally coupled to
gravity, with Lagrangian
\begin{equation}
{\cal L}=\frac12(D^\mu\Phi^{aa'})(D_\mu\Phi^{aa'})
-\frac{y}{6}f^{abc}\tilde{f}^{a'b'c'}
\Phi^{aa'}\Phi^{bb'}\Phi^{cc'}-\frac{\cal R}{12}\Phi^{aa'}\Phi^{aa'},
\label{biadjoint-conformal}
\end{equation}
where ${\cal R}$ is the Ricci scalar, a property making use of the
fact that ${\cal R}\propto \lambda$ for de Sitter spacetime. The
constant of proportionality is precisely such as to make
eq.~(\ref{TNPhieq}) follow from eq.~(\ref{biadjoint-conformal}), in
four spacetime dimensions.\\

Having reviewed all necessary details regarding the Kerr-Schild
double copy, we now turn to the investigation of other curved space
examples.

\section{Kerr-Schild solutions in curved space}
\label{sec:KS}

As stated in the introduction, our examination of curved space
instances of the classical double copy is motivated by the results of
ref.~\cite{Adamo:2017nia}, concerning a double copy of type A. This
associates a gauge theory amplitude in the presence of a non-trivial
background field, with a gravity amplitude defined with respect to a
non-Minkowski background metric, where the gauge and gravity
background fields should be related. In this section, we will see that
Kerr-Schild solutions indeed provide a natural framework for
constructing such double copies for exact field solutions, rather than
perturbative amplitudes.

\subsection{Single Kerr-Schild solutions}
\label{sec:singleKS}

The simplest such examples can be constructed, albeit rather
artificially, by starting with single Kerr-Schild metrics around
Minkowski space. We may split up such solutions according to
\begin{align}
g_{\mu\nu}&=\eta_{\mu\nu}+\phi k_\mu k_\nu\notag\\
&=\eta_{\mu\nu}+\phi_1 k_\mu k_\nu+\phi_2 k_\mu k_\nu,
\label{KSsplit}
\end{align}
where we have introduced
\begin{equation}
\phi_1=\xi\phi,\quad \phi_2=(1-\xi)\phi,\quad 0\leq\xi\leq 1.
\label{phi12def}
\end{equation}
Thus, as is well-known (see e.g.~\cite{Stephani:2003tm}), any given
single Kerr-Schild metric can always be thought of as a double
Kerr-Schild metric. Following the discussion of
section~\ref{sec:review}, it is straightforward to single copy
eq.~(\ref{KSsplit}) term-by-term, resulting in the gauge field
\begin{align}
A_\mu^a=c^a\Big[
\phi_1 k_\mu +\phi_2 k_\mu
\Big].
\label{KSsplitgauge}
\end{align}
This is itself a rewriting of eq.~(\ref{SingleCopy}), that is
ultimately possible due to the linearity of the field equations in the
Kerr-Schild double copy. However, we can reinterpret
eqs.~(\ref{KSsplit}, \ref{KSsplitgauge}) as follows. By defining
\begin{equation}
\bar{g}_{\mu\nu}=\eta_{\mu\nu}+\phi_1 k_\mu k_\nu,
\label{gbardef}
\end{equation}
we may rewrite eq.~(\ref{KSsplit}) as 
\begin{equation}
g_{\mu\nu}=\bar{g}_{\mu\nu}+\tilde{h}_{\mu\nu},\quad
\tilde{h}_{\mu\nu}=\phi_2 k_\mu k_\nu,
\label{KSsplit2}
\end{equation}
so that the solution of eq.~(\ref{KSsplit}) may be regarded as
containing a graviton field involving only the field $\phi_2$, defined
with respect to the non-Minkowski background $\bar{g}_{\mu\nu}$. On
the gauge theory side, we can define
\begin{equation}
\bar{A}_\mu^a=c^a \phi_1 k_\mu,
\label{KSsplit3}
\end{equation}
so that the solution of eq.~(\ref{KSsplitgauge}) becomes
\begin{equation}
A^a_\mu=\bar{A}_\mu^a+\tilde{A}_\mu^a,\quad
\tilde{A}_\mu^a=c^a\phi_2 k_\mu.
\label{Atildedef}
\end{equation}
This is thus our first example of a type A curved space double copy,
for classical solutions rather than amplitudes. A gauge field defined
with respect to a non-trivial background field copies to a graviton
field with a non-trivial background, where the two backgrounds are
themselves related (i.e. they are themselves Kerr-Schild, so we know
how to double copy them).\\

As indicated in figure~\ref{fig:curvedmap}, there is another way to
consider double copies in curved space (type B). Namely, it may be
possible to single copy a graviton defined with respect to a
non-Minkowski background, to a gauge field living on the same
background. To this end, one may consider the graviton field
$\tilde{h}_{\mu\nu}$ of eq.~(\ref{KSsplit2}), which single copies to
the field $\tilde{A}^a_\mu$ of eq.~(\ref{Atildedef}). On the gauge
theory side, one may impose the same background $\bar{g}_{\mu\nu}$,
and examine the curved space Maxwell equations
\begin{equation}
D^\mu \tilde{F}_{\mu\nu}^a=j_\nu,
\label{KSsplitMaxwell}
\end{equation}
where $\tilde{F}_{\mu\nu}^a$ is the field strength tensor formed from
the gauge field $\tilde{A}^a_\mu$. For a consistent double copy of
type B, one requires that the source current is somehow related to
the energy-momentum tensor in a recognisable way, so that the two
solutions are related. Let us give two examples. Firstly, one may
consider the Schwarzschild metric, for which
\begin{equation}
\phi=\frac{2M}{r},\quad k^\mu=(1,1,0,0),
\label{Schwarzschild}
\end{equation}
where we adopt spherical polar coordinates
$(t,r,\theta,\phi)$. Writing the graviton as
\begin{equation}
h_{\mu\nu}=\frac{2M_1}{r}k_\mu k_\nu+\frac{2M_2}{r}k_\mu k_\nu,\quad
M_1+M_2=M,
\label{KSsplitSW}
\end{equation}
we may define the background field
\begin{equation}
\bar{g}_{\mu\nu}=\eta_{\mu\nu}+\bar{h}_{\mu\nu},\quad
\bar{h}_{\mu\nu}=\frac{2M_1}{r}k_\mu k_\nu,
\label{gbarSW}
\end{equation}
and then single copy the graviton
\begin{equation}
\tilde{h}_{\mu\nu}=\frac{2M_2}{r}k_\mu k_\nu
\label{tildehSW}
\end{equation}
to get a gauge field
\begin{equation}
\tilde{A}_\mu^a=\frac{c^a}{r} k_\mu.
\label{AtildeSW}
\end{equation}
The curved space Maxwell equations of eq.~(\ref{KSsplitMaxwell}) then
yield~\footnote{Here, we do not include the delta function source at
  the origin, corresponding to the point charge (mass) sourcing the
  field $\tilde{A}^a_\mu$ ($h_{\mu\nu}$).}
\begin{equation}
j_\mu^a=0,
\label{jmuSW}
\end{equation}
which is indeed consistent: the Schwarzschild metric is a vacuum
solution in General Relativity. Here we find that its curved space
single copy is also a (gauge theory) vacuum solution, on the curved
space defined by $\bar{g}_{\mu\nu}$. \\

A second example is given by de Sitter spacetime, which has the
Kerr-Schild form
\begin{equation}
\phi=\lambda r^2,\quad k_\mu=(-1,1,0,0),
\label{dS}
\end{equation}
where $\lambda$ is the cosmological constant. Splitting this similarly
to eq.~(\ref{KSsplitSW}) gives
\begin{equation}
h_{\mu\nu}=\lambda_1 r^2 k_\mu k_\nu+\lambda_2 r^2 k_\mu k_\nu,\quad
\lambda_1+\lambda_2=\lambda.
\label{KSsplitdS}
\end{equation}
We can then define the graviton
\begin{equation}
\tilde{h}_{\mu\nu}=\lambda_2 r^2 k_\mu k_\nu,
\label{tildehdS}
\end{equation}
whose single copy gauge field
\begin{equation}
\tilde{A}^a_\mu=c^a \lambda_2 r^2 k_\mu
\label{AtildedS}
\end{equation}
satisfies the curved space Maxwell equation with
\begin{equation}
j^a_\nu=(6\lambda_2,0,0,0).
\label{jmudS}
\end{equation}
Again this makes sense: the graviton is sourced by a constant energy
density filling all space which, in the gauge theory, becomes a
constant charge density. The single copy has thus turned momentum
degrees of freedom into colour degrees of freedom, precisely as in the
flat space case examined in ref.~\cite{Luna:2015paa} (and also the
single copy for
amplitudes~\cite{Bern:2008qj,Bern:2010ue,Bern:2010yg}). \\

We have not been able to prove in general that the curved space
Maxwell equations are satisfied for arbitrary single Kerr-Schild
solutions that are rewritten in the form of
eq.~(\ref{KSsplit2}). However, we have at least shown for some special
-- and, indeed, astrophysically relevant -- cases, a type B double
copy map is possible. The question then arises of how general this map
is. The conventional double copy, in its simplest form, relates a
gauge theory to a gravity theory. A gauge theory on a curved
background (even if this is non-dynamical) would appear to involve
gravity, and thus this type of double copy map seems to relate a
coupled Einstein-gauge theory system to itself. One does not then
expect this map to be fully general, or to apply to supersymmetric
generalisations that are known to work in flat space. \\

Evidence towards this viewpoint can be gleaned by examining the zeroth
copy. As discussed in section~\ref{sec:review}, the Kerr-Schild field
$\phi$ is found to satisfy the linearised biadjoint scalar field
equation, and can be interpreted as a scalar propagator. In the type
II double copy, we can take the zeroth copy of the gauge field
$\tilde{A}_\mu^a$ to generate a scalar field
\begin{equation}
\tilde{\Phi}^{aa'}=c^a\tilde{c}^{a'}\phi_2,
\label{Phizeroth}
\end{equation}
and consider the curved space linearised biadjoint equation
\begin{equation}
D^\mu D_\mu \Phi^{aa'}=c^a\tilde{c}^{a'}\xi,
\label{biadjointcurved}
\end{equation}
which defines $\xi$. For the Schwarzschild and de Sitter examples, we
find
\begin{equation}
\xi_{\rm SWC}=-\frac{4M_1 M_2}{r^4},\quad 
\xi_{\rm dS}=6\lambda_2-10r^2\lambda_1\lambda_2
\label{xiaa'}
\end{equation}
respectively, which we can not straightforwardly interpret as being
related to the source current in the gauge theory. It thus seems that
the type B double copy can indeed associate a gauge theory solution
in curved space with a gravity counterpart, at the expense of not
having a consistent zeroth copy. This also sheds light on the
speculation of ref.~\cite{Luna:2015paa}, that the zeroth copy for a
curved background may result in a biadjoint scalar theory conformally
coupled to gravity (eq.~(\ref{biadjoint-conformal})). The results of
eq.~(\ref{xiaa'}) provide a simple counter-example to this conjecture,
showing that the situation is more complex than previously thought.

\subsection{Multiple Kerr-Schild solutions}
\label{sec:multiKS}

In the previous section, we used single Kerr-Schild solutions to
provide some first examples of curved space double copies, of both
type A and type B. Here, we study whether such conclusions also apply
to more complicated solutions. As a first generalisation, we may
consider multiple Kerr-Schild solutions in Minkowski space, namely
those of form
\begin{equation}
g_{\mu\nu}=\eta_{\mu\nu}+\sum_i \phi_i k^{(i)}_\mu k^{(i)}_\nu,
\label{multiKS}
\end{equation}
where each vector $k^{(i)}_\mu$ is null and geodetic with respect to
both the Minkowski and full metric, and the set of Kerr-Schild vectors
obeys the mutual orthogonality relations
\begin{equation}
\eta^{\mu\nu} k^{(i)}_{\mu} k^{(j)}_{\nu}=
g^{\mu\nu} k^{(i)}_{\mu} k^{(j)}_{\nu}=0,\quad \forall i, j.
\label{kikj}
\end{equation}
In certain cases, this ansatz linearises the mixed Ricci tensor
$R^\mu_\nu$, and thus provides an exact solution of the Einstein
equations~\footnote{Examples involving more than two terms are the
  higher dimensional Taub-NUT-like solutions of
  refs.~\cite{Chen:2006xh,Chen:2007fs}.}. We further consider the
general class of multi-Kerr-Schild solutions in which each term in the
graviton is itself a solution of the linearised Einstein equations. In
the stationary case, we may then single copy eq.~(\ref{multiKS}) to
produce a gauge field
\begin{equation}
A^a_\mu=c^a \sum_i \phi_i k^{(i)}_\mu.
\label{multiKSgauge}
\end{equation}
Given that each term in the graviton constitutes a stationary Kerr-Schild
solution, the results of ref.~\cite{Monteiro:2014cda} immediately
imply that each term in eq.~(\ref{multiKSgauge}) satisfies the
linearised Yang-Mills equations. Linearity then implies that the
complete field of eq.~(\ref{multiKSgauge}) is also a solution, and
thus a well-defined single copy of the gravity result. \\

As for the solution of eq.~(\ref{KSsplit}), we can use any
multi-Kerr-Schild solution of the form of eqs.~(\ref{multiKS},
\ref{multiKSgauge}) to construct a type A curved space double copy. To
do this, one may partition the terms in eq.~(\ref{multiKS}) into two
sets $\Gamma_1$ and $\Gamma_2$, before defining
\begin{equation}
\bar{g}_{\mu\nu}=\eta_{\mu\nu}+\sum_{i\in \Gamma_1}\phi_ik^{(i)}_\mu
k^{(i)}_\nu,\quad \bar{A}^a_\mu=c^a\sum_{i\in \Gamma_1}\phi_ik^{(i)}_\mu,
\label{multiKSsplit}
\end{equation}
and
\begin{equation}
\tilde{h}_{\mu\nu}=\eta_{\mu\nu}+\sum_{i\in \Gamma_2}\phi_ik^{(i)}_\mu
k^{(i)}_\nu,\quad \tilde{A}^a_\mu=c^a\sum_{i\in \Gamma_2}\phi_ik^{(i)}_\mu.
\label{AtildemultiKS}
\end{equation}
The full gravity and gauge fields may now be written as
\begin{equation}
g_{\mu\nu}=\bar{g}_{\mu\nu}+\tilde{h}_{\mu\nu},\quad
A_\mu^a=\bar{A}_\mu^a+\tilde{A}_\mu^a.
\label{AfullmultiKS}
\end{equation}
This is indeed an example of the type A double copy shown in
figure~\ref{fig:curvedmap}: the gauge field $\tilde{A}^a_\mu$ defined
with respect to the background field $\bar{A}_\mu^a$ double copies to
the graviton $\tilde{h}_{\mu\nu}$, defined with respect to the
background metric $\bar{g}_{\mu\nu}$. \\

\begin{figure}
\begin{center}
\scalebox{0.8}{\includegraphics{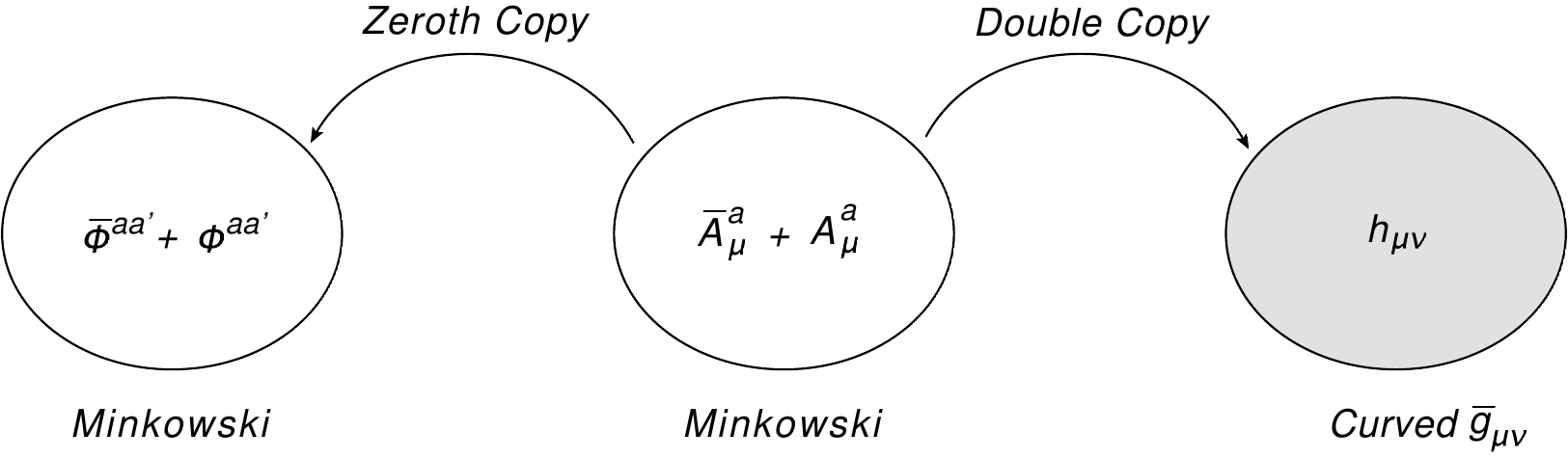}}
\caption{Generalisation of the type A double copy of
  figure~\ref{fig:curvedmap} to include the zeroth copy, which relates
  the gauge field defined with a non-trivial background to similar
  solutions in a biadjoint scalar theory.}
\label{fig:curvedmap2}
\end{center}
\end{figure}

Furthermore, the zeroth copy is also well-defined, as for the flat
space classical double copy: from eq.~(\ref{multiKSgauge}), we may
define the biadjoint field
\begin{equation}
\Phi^{aa'}=c^a\tilde{c}^{a'}\sum_i \phi_i.
\label{biadjointmultiKS}
\end{equation}
The fact that each term in the gauge field satisfies the linearised
Yang-Mills equations implies, again from ref.~\cite{Monteiro:2014cda},
that each term in eq.~(\ref{biadjointmultiKS}) satisfies the
linearised biadjoint scalar equation. Similarly to
eq.~(\ref{multiKSsplit}), we may then define
\begin{equation}
\bar{\Phi}^{aa'}=c^a\tilde{c}^{a'}\sum_{i\in \Gamma_1}\phi_i,\quad
\tilde{\Phi}^{aa'}=c^a\tilde{c}^{a'}\sum_{i\in \Gamma_2}\phi_i,
\label{PhisplitmultiKS}
\end{equation}
so that the full biadjoint field can be written
\begin{equation}
\Phi^{aa'}=\bar{\Phi}^{aa'}+\tilde{\Phi}^{aa'}.
\label{PhisplitmultiKS2}
\end{equation}
This is a direct analogue of the type A double copy between gauge
theory and gravity: a classical field defined with respect to a
background copies between biadjoint scalar and gauge theory. The
relationship between the three theories is shown in
figure~\ref{fig:curvedmap2}. Given that we will always be talking
about solutions of the linearised Yang-Mills equations from now on, we
will omit colour indices and vectors in what follows.\\

We may also examine whether or not it is possible to construct a type
B double copy for multi-Kerr-Schild solutions, by considering
specific examples. In section~\ref{sec:singleKS}, we saw that this was
indeed possible for the Schwarzschild and de Sitter solutions, split
according to eqs.~(\ref{KSsplit}, \ref{phi12def}). More generally, we
can take either of these gravitons as part of the background metric
$\bar{g}_{\mu\nu}$, and allow either of them to be the perturbation
$\tilde{h}_{\mu\nu}$. The full list of possibilities is enumerated in
table~\ref{tab:doubleKS}, where the full metric is given by
eq.~(\ref{KSsplit}), with $k^\mu=(1,1,0,0)$ in spherical polar
coordinates. The first two rows contain the pure Schwarzschild (SWC)
and de Sitter (dS) metrics, and the third / fourth rows the cases
already considered in the previous section. Finally, the fifth and
sixth rows contain the metric formed by perturbing the Schwarzschild
solution with a de Sitter Kerr-Schild graviton, and vice versa. For
each metric, we give an expression for the timelike component $j^t$ of
the source current that appears in the curved space Maxwell equation
of eq.~(\ref{KSsplitMaxwell}) (the spacelike components are found to
vanish in all cases), as well as the quantity $\xi$ that appears on
the right-hand side of the curved space linearised biadjoint equation
(eq.~(\ref{biadjointcurved})).\\
\begin{table}
\begin{center}
\begin{tabular}{|c |c |c |c |c|}
\hline
Metric&$\phi_1$&$\phi_2$&$j^t$&$\xi$\\ \hline
SWC&$0$&$2m_2/r$&$0$&$0$\\ \hline
dS&$0$&$\lambda_2 r^2$&$6\lambda_2$&$6\lambda_2$\\ \hline
SWC+SWC&$2m_1/r$&$2m_2/r$&$0$&$-4m_1m_2/r^4$
\\ \hline
dS+dS&$\lambda_1 r^2$&$\lambda_2 r^2$&$6\lambda_2$&$6\lambda_2-10r^2 \lambda_1 \lambda_2$\\ \hline
SWC+dS&$2m_1/r$&$\lambda_2 r^2$&$6\lambda_2$&$6\lambda_2-8m_1\lambda_2/r$\\ \hline
dS+SWC&$\lambda_1 r^2$&$2m_2/r$&$0$&$4\lambda_1m_2/r$\\ \hline
\end{tabular}
\caption{Table of type B single and zeroth copies of Kerr-Schild
  metrics of the form of eq.~(\ref{KSsplit}), where $\phi_1$ and
  $\phi_2$ are allowed to be different. Here A+B denotes a Kerr-Schild
  graviton for metric B considered as a perturbation on background
  metric A, where SWC and dS represent the Schwarzschild and de Sitter
  gravitons respectively.}
\label{tab:doubleKS}
\end{center}
\end{table}

In all cases, the type B single copy indeed holds. That is, the gauge
theory contains a source current consistent with the perturbation term
in the gauge field: zero in the Schwarzschild case~\footnote{As
  earlier, we do not bother showing the delta function source at the
  origin.}, and a uniform charge density in the de Sitter case, whose
counterpart in gravity is the cosmological constant. There are no
terms in the source current which are sensitive to the field $\phi_1$,
which would invalidate the picture of figure~\ref{fig:curvedmap}. The
zeroth copy holds only in the cases of a pure single Kerr-Schild
solution (i.e. the cases considered in the original classical double
copy of refs.~\cite{Monteiro:2014cda,Luna:2015paa}). For all of the
double Kerr-Schild solutions, the source includes a position-dependent
term that has no immediately evident counterpart in the gauge or
gravity theory. \\

In the above examples, the full metric contains two Kerr-Schild terms,
each of which has the same vector $k^\mu$, corresponding to a
spherically symmetric system. We can then ask what the most general
results for $j^t$ and $\xi$ are, for unspecified functions $\phi_1(r)$
and $\phi_2(r)$. The results are
\begin{equation}
j^t=\frac{2\phi_2'(r)}{r}+\phi_2''(r)=\nabla_{\rm M}^2 \phi_2,\quad
\xi=\nabla^2 \phi_2=
j^t(1-\phi_1(r))-\phi'_1(r)\phi'_2(r). 
\label{jtxi}
\end{equation}
Here $\nabla^2$ is the Laplacian operator associated with the full
background metric, and $\nabla^2_{\rm M}$ the corresponding operator
in Minkowski space. We thus conclude that if $\phi_2$ is associated
with a vacuum solution in Minkowski space, the type II single copy is
well-defined, in that it is also a vacuum solution. However, the
source for the zeroth copy involves the background field $\phi_1$ and
thus does not seem to have a meaningful interpretation. Of course, the
fields $\phi_1$ and $\phi_2$ in eq.~(\ref{jtxi}) are not arbitrary,
but must be fixed by the Einstein equations. For the case of
spherically symmetric (and stationary) vacuum solutions up to the
presence of a cosmological constant, the only possible solutions are
the Schwarzschild and de Sitter cases examined already in
table~\ref{tab:doubleKS}. Nevertheless, the general form of the
current in eq.~(\ref{jtxi}) does not rule out that there may be
non-trivial solutions with extended sources, such that one may still
find a consistent single copy interpretation. It is not known even in
the flat space case how to construct such maps (see
e.g. refs.~\cite{Ridgway:2015fdl,Luna:2016due} for discussions of
source terms in various contexts).\\

Above we have discussed cases in which the background scalar field
$\phi_1$ is spherically symmetric. Our results are more general than
this, however. We have explicitly checked that our conclusion that the
type B single copy is a vacuum solution if $\phi_2$ is associated with
a vacuum solution in Minkowski space, holds true even if $\phi_1$ has
an arbitrary spatial and temporal dependence. \\

It is furthermore useful to note that, as in the flat space cases
considered in ref.~\cite{Monteiro:2014cda}, one may transform the type
B single copy gauge field into a more recognisable form. Starting with
the gauge field in spherical polar coordinates,
\begin{displaymath}
A_\mu=\phi_2(-1,1,0,0),
\end{displaymath}
one may perform a gauge transformation
\begin{equation}
A_\mu\rightarrow A'_\mu=A_\mu+D_\mu \chi(r)=A_\mu+\partial_\mu \chi(r),
\label{AmuGT}
\end{equation}
where
\begin{equation}
\chi(r)=-\int^r dr'\,\phi_2(r'),
\label{chidef}
\end{equation}
so that eq.~(\ref{AmuGT}) implies
\begin{equation}
A'_\mu=(-\phi_2,0,0,0).
\label{A'def}
\end{equation}
Thus, $\phi_2$ indeed has the interpretation of an electrostatic
potential.\\

As implied above by the above results, the type B double copy is not
necessarily expected to be a fully general map between exact solutions
in gauge and gravity theories in curved space. However, it is
interesting to examine whether or not it shares the property of the
type A (and amplitude) double copies, in being independent of the
number of spacetime dimensions $d$. Indeed, one may show that for a
$d$-dimensional background metric $\bar{g}_{\mu\nu}$ of the form of
eq.~(\ref{gbardef}), the gauge field $\tilde{A}_\mu$ of
eq.~(\ref{Atildedef}) satisfies the Maxwell equations, with a current
density given by~\footnote{Equation~(\ref{currD}) also turns out to be
  true when the field $\phi_1$ depends on time and the non-radial
  spatial coordinates.}
\begin{equation}
j^\mu=(\nabla_{\rm M}^2 \phi_2,0,0\ldots,0),
\label{currD}
\end{equation}
where the Minkowski-space Laplacian on the right-hand side is in
$(d-1)$ space dimensions. Thus, our above discussion generalises for
any $d$. We present a proof of these statements in
appendix~\ref{app:Ddim}. \\

Having examined multiple Kerr-Schild solutions where each term
contains the same Kerr-Schild vector $k^\mu$, it is instructive to
instead consider an example in which these vectors can be different.
One such example is the Taub-NUT solution, for which the metric takes
the form
\begin{eqnarray}
g_{\mu\nu}=\eta_{\mu\nu}+\phi k_\mu k_\nu+\psi l_\mu l_\nu.
\end{eqnarray}
The Minkowski line element can be written as
\begin{eqnarray}
ds^2=-dt^2+\frac{\rho^2}{a^2+r^2}dr^2+\rho^2d\theta^2+(r^2+a^2)\sin^2\theta
d\varphi^2
\label{Minkspheroidal}
\end{eqnarray}
in spheroidal coordinates, where 
\begin{eqnarray}
\rho^2\equiv r^2+a^2\cos^2\theta.
\label{rhodef}
\end{eqnarray}
The vectors $k_\mu$ and $l_\mu$ are defined by 
\begin{eqnarray}
l_\mu dx^\mu=dt+\frac{\rho^2}{a^2+r^2}dr-a\sin^2\theta d\varphi\\
k_\mu dx^\mu=dt-\frac{i\rho^2}{a\sin\theta}d\theta+\frac{r^2+a^2}{a}d\varphi,
\label{kldef}
\end{eqnarray}
while the scalar functions $\phi$ and $\psi$ are given by
\begin{eqnarray}
\psi=\frac{2mr}{\rho^2},\qquad \phi=\frac{2la\cos\theta}{\rho^2}.
\label{psiphidef}
\end{eqnarray}
As for the various metrics considered in table~\ref{tab:doubleKS}, in
considering the type B single copy, we can take either of the
Kerr-Schild terms to be part of the background metric, resulting in
two possibilities:
\begin{align*}
\textrm{Case 1:}\qquad g_{\mu\nu}=\bar{g}_{\mu\nu}+\phi k_\mu k_\nu,\qquad \bar{g}_{\mu\nu}=\eta_{\mu\nu}+\psi l_\mu l_\nu,\\
\textrm{Case 2:}\qquad g_{\mu\nu}=\bar{g}_{\mu\nu}+\psi l_\mu l_\nu,\qquad \bar{g}_{\mu\nu}=\eta_{\mu\nu}+\phi k_\mu k_\nu.
\end{align*}
The gauge fields obtained from the single copy of the perturbation
term in both cases satisfy homogeneous Maxwell equations
\begin{eqnarray}
j^\nu=0,
\end{eqnarray}
so that the single copy is indeed consistent (n.b. the Taub-NUT
solution is a vacuum solution). The zeroth copy factor is given in
both cases by 
\begin{eqnarray}
\xi=\frac{4\phi \psi (\rho^2-2r^2)}{\rho^4},
\label{xiTN}
\end{eqnarray}
so that, consistently with our previous results, the type B single
copy does not appear to be meaningful.\\

Let us summarise the results of this section. We have examined whether
it is possible to construct a double copy for classical solutions that
mimics the result found for amplitudes in
ref.~\cite{Adamo:2017nia}. We have indeed found such a procedure,
based on the same Kerr-Schild solutions that were used to formulate a
flat space double copy in
refs.~\cite{Monteiro:2014cda,Luna:2015paa,Luna:2016due}. In this
picture, a gauge field defined with respect to a non-trivial
background field copies to a graviton defined with respect to a
background metric, where the background fields in the two theories are
related. We were able to relate the background fields due to the fact
that they obeyed the original Kerr-Schild double copy by
themselves. Furthermore, there is a well-defined zeroth copy, in which
the resulting biadjoint field is also defined with respect to a
background, where the latter is the zeroth copy of the background
gauge field. We call this procedure a {\it type A curved space double
  copy}, to distinguish it from an alternative procedure ({\it type
  B}) in which the gauge field lives on a non-dynamical curved
spacetime, and copies to a graviton field defined with respect to the
same spacetime. In this picture, the zeroth copy does not appear to be
meaningful, in that the biadjoint field appears not to be physically
related to its gauge theory counterpart, due to the presence of
unwanted source terms.\\

In all of the above cases, we knew how to construct a type A double
copy due to the fact that we could relate the background gauge field
with its gravitational counterpart. The type B double copy, however,
does not require such a relationship, as the same curved metric
appears in both the gauge and gravity theories. It is then interesting
to look for examples of this relationship in which the background
metric is {\it not} of Kerr-Schild form, and thus cannot be
single-copied according to the procedure of
refs.~\cite{Monteiro:2014cda,Luna:2015paa,Luna:2016due}. We have
indeed found such examples, which we describe in the following
section.

\section{Conformally flat background metrics}
\label{sec:Einstein}

In this section, we consider conformally flat spacetimes. More
specifically, we consider spacetimes whose metrics can be written (in
some coordinate system) as a conformal transformation of Minkowski
space:
\begin{equation}
\bar{g}_{\mu\nu}=\Omega^2(x^\mu) \eta_{\mu\nu}.
\label{conformflat}
\end{equation}
As the bar notation on the left-hand side already suggests, we will
use such metrics as background metrics for Kerr-Schild solutions. This
will work for any conformally flat metric, given that if $k^\mu$ is
null and geodetic with respect to the Minkowski metric, it is
straightforward to show that it is also null and geodetic with respect
to $\bar{g}_{\mu\nu}$. \\

As a warm-up, let us examine the case where the background 
is the well-known Einstein
static universe. For convenience, we will adopt the coordinates
and conventions of ref.~\cite{1981AnPhy.134..326T}, such that the line
element is
\begin{eqnarray}
d\bar{s}^2=-dt^2+dr^2-2a\sin^2\theta d\varphi dr+\frac{|\beta|^2}{\mathcal{D}^2}d\theta^2+(|\beta|^2+a^2\sin^2\theta)\sin^2\theta d\varphi^2,
\label{ds2Einstein}
\end{eqnarray}
where
\begin{align*}
\mathcal{D}&=1-(a^2/R_0^2)\sin^2\theta,\\
\beta&=(R_0^2-a^2)^{1/2}\sin\frac{r}{R_0}+ia\cos{\theta}.
\end{align*}
The Ricci tensor and scalar for this metric take
the particularly simple forms
\begin{eqnarray}
\bar{R}_{\mu\nu}=\frac{2}{R_0^2}(\bar{g}_{\mu\nu}+\bar{u}_\mu\bar{u}_\nu),\qquad \bar{R}=\frac{6}{R_0^2},
\label{RicciEinstein}
\end{eqnarray}
respectively, where $u_\mu$ is the unit timelike vector given by
\begin{eqnarray}
\bar{u}_\mu=(1,0,0,0),\qquad \bar{u}^\mu=(-1,0,0,0).
\label{udef}
\end{eqnarray}
We can construct a solution
\begin{eqnarray}
g_{\mu\nu}=\bar{g}_{\mu\nu}+2H k_\mu k_\nu
\label{EinsteinKS}
\end{eqnarray}
of single Kerr-Schild form, where for the Kerr-Schild term we adopt
the notation of ref.~\cite{1981AnPhy.134..326T} for ease of
comparison. 
The Kerr-Schild vector $k_\mu$ is defined by
\begin{eqnarray}
\sqrt{2}k_\mu=(-1,-1,0,a\sin^2\theta),
\label{Einsteink}
\end{eqnarray}
while the scalar function
\begin{eqnarray}
H=m\cdf_\mu k^\mu,
\label{EinsteinH}
\end{eqnarray}
with $\cdf_\mu$ the covariant derivative associated with
$\bar{g}_{\mu\nu}$. The solution defined by
eqs. (\ref{EinsteinKS}-\ref{EinsteinH}) corresponds to a rotating
black hole over the Einstein static universe. In order to further
examine the effect of this perturbation, we note that the mixed-index
Ricci tensor of the full metric can be recast in the form
\begin{eqnarray}
R^\mu_{\ \nu}=-\frac{2}{R_0^2}(1-H)(\delta^\mu_\nu+u^\mu u_\nu),
\label{RmunuEinstein}
\end{eqnarray}
where we have introduced the vectors
\begin{eqnarray}
u^\mu=\frac{\bar{u}^\mu}{\sqrt{1-H}},\qquad u_\mu=\frac{1}{\sqrt{1-H}}(\bar{u}_\mu+\sqrt{2}Hk_\mu).
\label{udef2}
\end{eqnarray}
The Einstein equations then become
\begin{eqnarray}
R^\mu_{\ \nu}-\frac{1}{2}\delta^\mu_\nu R=-8\pi
((\rho+p)u^\mu u_\nu+p\delta^\mu_\nu)+\Lambda \delta^\mu_{\ \nu}.
\label{Einstein_eqs}
\end{eqnarray} 
That is, the matter content of the theory is that of a perfect fluid,
whose energy density $\rho$ and pressure $p$ are given in this case by
\begin{align}
8\pi\rho&=\frac{3}{R_0^2}(1-H)-\Lambda,\\
8\pi p&=-\frac{1}{R_0^2}(1-H)+\Lambda.
\label{rhopdef}
\end{align}
We see that the presence of the Kerr-Schild term acts to redefine the
parameters associated with the background metric, reminiscent of the
split Kerr-Schild metrics we considered in
section~\ref{sec:singleKS}. A number of other such solutions are
presented in ref.~\cite{1981AnPhy.134..326T}. \\

We may single copy the graviton appearing in eq.~(\ref{EinsteinKS}) by
defining the gauge field
\begin{equation}
A_\mu^a=c^a H k_\mu,
\label{AmuEinstein}
\end{equation}
which we find satisfies the homogeneous linearised Yang-Mills equation
\begin{equation}
D_\mu F^{\mu\nu}=0,
\label{YMEinstein}
\end{equation}
where $D_\mu$, as above, is the covariant derivative for the Einstein
static universe. This provides an example of the type B double copy
of figure~\ref{fig:curvedmap}: on the gravity side, a fluid is needed
to source the background metric. There is no corresponding source
current in the gauge theory, as there is no background gauge field,
unlike in the type A double copy.  \\

In the previous examples of the type B double copy, we saw that the
zeroth copy did not appear to have a meaningful
interpretation. Interestingly, in the present example the field $H$
satisfies the homogeneous linearised biadjoint scalar equation
\begin{equation}
D^2 H=0,
\label{biadjointEinstein}
\end{equation}
which indeed leads to a well-defined zeroth copy for this case.\\

Having seen a particular example of the type B single copy for
non-Kerr-Schild backgrounds, let us now consider the general case of
background metrics of the form of eq.~(\ref{conformflat}), where the
Minkowski metric is given in spherical polar coordinates
$(t,r,\theta,\phi)$, so that the conformally transformed metric takes
the form
\begin{equation}
\bar{g}_{\mu\nu}=\Omega^2(x^\mu)
{\rm diag}(-1,1,r^2,r^2\sin^2\theta),
\label{gbarconformal}
\end{equation} 
Upon constructing the gauge field
\begin{equation}
A_\mu=k_\mu \phi_2(r),\quad k_\mu=(-1,1,0,0),
\label{Amuconformal}
\end{equation}
we find that this satisfies the curved space Maxwell equation (in the
spacetime whose metric is $\bar{g}_{\mu\nu}$)
\begin{equation}
D_\mu F^{\mu\nu}=j^\nu,\quad j^\nu=\left(\frac{\nabla_{\rm M}^2\phi_2}
{\Omega^4(x^\mu)},0,0,0
\right),
\label{Maxwellconformal}
\end{equation}
where $\nabla^2_{\rm M}$ is the Minkowski space Laplacian
operator. Note that this result does not require the conformal factor
$\Omega$ to have spherical symmetry - it may be a general function of
$(t,r,\theta,\phi)$. From eq.~(\ref{Maxwellconformal}), we see that if
the gauge field of eq.~(\ref{Amuconformal}) satisfies a vacuum Maxwell
equation in Minkowski space, it also does so in the conformally
transformed metric, analogous to the double Kerr-Schild examples
considered in the previous section. Thus, the Minkowski space single
copy extends to a type B curved space double copy, even though the
background metric $\bar{g}_{\mu\nu}$ does not have a Kerr-Schild form,
and thus is not immediately amenable to a type A single copy. We may
also examine the type B zeroth copy, and one finds the curved space
linearised biadjoint equation
\begin{equation}
D^\mu D_\mu \phi_2=\frac{\nabla^2_{\rm M}\phi_2}{\Omega^2(x^\mu)}
+\frac{2\phi_2'(r)\partial_r\Omega(x^\mu)}{\Omega^3(x^\mu)}.
\label{biadjointconformal}
\end{equation}
The second term on the right-hand side involves a spatial derivative
of the conformal factor, which is not present in the gauge theory
source. Thus, it does not seem possible to interpret the zeroth copy
in general, in line with our previous conclusions for the type B
procedure.

\section{Discussion}
\label{sec:discuss}

The aim of this paper has been to examine whether or not the classical
double copy of refs.~\cite{Monteiro:2014cda,Luna:2015paa,Luna:2016due}
can yield gravitons defined around a non-Minkowski background
metric. Our motivation is the recent study of~\cite{Adamo:2017nia},
which constructs such a procedure for amplitudes. They consider gauge
fields corresponding to perturbations around a plane-wave solution,
whose amplitudes double copy to amplitudes for gravitons defined with
respect to a gravitational plane wave background. We call this an
example of a {\it type A} curved space double copy, depicted
schematically in figure~\ref{fig:curvedmap}, and in which a gauge
field defined with respect to a non-trivial background copies to a
graviton defined with respect to a non-Minkowsi metric, where the
background fields in both cases are related. We have shown that one
can indeed construct such a double copy for classical solutions, based
on Kerr-Schild solutions, which underlie the classical double copy of
refs.~\cite{Monteiro:2014cda,Luna:2015paa,Luna:2016due} in flat
space. Furthermore, there is a well-defined zeroth copy, which maps
the gauge field to a biadjoint scalar field, satisfying a linearised
equation of motion. This is itself interesting for the curved space
amplitude double copy of ref.~\cite{Adamo:2017nia}. There, the authors
considered three-point amplitudes, which do not contain propagator
factors. For the flat space double copy, the fact that the zeroth copy
exists for classical solutions is related to how one deals with
propagator factors in amplitudes. Thus, the fact that the zeroth copy
also works for the type A curved space double copy suggests that the
results of ref.~\cite{Adamo:2017nia} can indeed be generalised to
higher-point amplitudes. \\

We also saw that it was possible to interpret the type A classical
double copy in an alternative way, namely that one can associate a
graviton defined with respect to a non-Minkowski background metric,
with a gauge field living on the same spacetime. We named this a {\it
  type B} double copy, and presented a number of non-trivial
examples. In almost all of the cases studied here, the zeroth copy
does not have a meaningful interpretation, suggesting that the type B
double copy is not a fully general relationship between gauge and
gravity theories that is rooted in first principles, but rather a map
that applies in certain special cases. Nevertheless, it could be very
useful to have such a map, particularly when the background spacetimes
(e.g. the de Sitter metric) are cosmologically relevant. Furthermore,
a type B double copy may exist even when it is not known how to
formulate a type A copy, due to e.g. having a background metric that
is not of Kerr-Schild form - we have here given the explicit example
of conformally flat metrics, including the Einstein static
universe. One may prove in general that for stationary spherically
symmetric gauge fields on a conformally flat metric, a vacuum Maxwell
equation in Minkowski space implies a vacuum solution on the curved
space, and thus a meaningful type B single copy. \\

Our results provide a way of extending the classical double copy of
ref.~\cite{Monteiro:2014cda,Luna:2015paa,Luna:2016due}, and will prove
useful in further investigations of the double copy, and its
applications. A number of avenues for further work suggest
themselves. Firstly, it would be interesting to know whether a type A
double copy can be set up for background metrics that are not
Kerr-Schild. Secondly, it would be useful to systematically determine
the circumstances in which the type B double copy applies, including
the possible addition of non-trivial source terms. Finally, one may
investigate whether the type A or type B double copies allow for new
insights or calculations relevant for astrophysics and cosmology.

\section*{Acknowledgments}

We thank Tim Adamo, Ricardo Monteiro and Donal O'Connell for
discussions and / or comments on the manuscript. CDW and NBA are
supported by the UK Science and Technology Facilities Council (STFC).
AL is funded by a Conacyt studentship, and thanks Queen Mary
University of London for hospitality. 

\appendix

\section{The type B single copy in $d$ dimensions}
\label{app:Ddim}

In this appendix, we consider the Maxwell equation for the gauge field
\begin{displaymath}
A_\mu=\phi_2(r) k_\mu,\quad k_\mu=(-1,1,0,\ldots,0)
\end{displaymath}
in a $d$-dimensional spacetime with spherical polar coordinates
$(t,r,\theta_1,\ldots, \theta_{d-2})$ whose metric is given by
\begin{displaymath}
\bar{g}_{\mu\nu}=\eta_{\mu\nu}+\phi_1(x^\rho) k_\mu k_\nu.
\end{displaymath}
Note that, for the sake of generality, we allow $\phi_1$ to depend on
all coordinates. One may then construct the field strength tensor
\begin{displaymath}
F_{\mu\nu}=D_\mu A_\nu-D_\mu A_\nu=\partial_\mu A_\nu-\partial_\nu A_\mu,
\end{displaymath}
where the second equality follows in the absence of torsion. It is
straightforward to show that the only non-zero components of this
tensor are
\begin{equation}
F_{tr}=-F_{rt}=-\phi'_2(r).
\label{Ftr}
\end{equation}
The curved space Maxwell equation is given by
\begin{equation}
D_\mu F^{\mu\nu}=\partial_\mu F^{\mu\nu}
+\bar{\Gamma}^\mu_{\mu\alpha}\,F^{\alpha \nu}
+\bar{\Gamma}_{\mu\alpha}^{\nu}\,F^{\mu\alpha}=j^\nu,
\label{Maxwellapp}
\end{equation}
where $\bar{\Gamma}^\alpha_{\beta\gamma}$ is the Christoffel symbol
associated with $\bar{g}_{\mu\nu}$. From the standard result
\begin{displaymath}
\bar{\Gamma}^{\alpha}_{\beta\gamma}=\frac12
\bar{g}^{\alpha\sigma}(\partial_\beta\bar{g}_{\gamma\sigma}
+\partial_\gamma\bar{g}_{\beta\sigma}
-\partial_\sigma\bar{g}_{\beta\gamma}),
\end{displaymath}
one finds that the only non-zero Christoffel symbols are given by
\begin{align}
\bar{\Gamma}^t_{tr}&=-\bar{\Gamma}^r_{rr}=
-\frac12(1+\phi_1)\partial_r\phi_1+\frac12
\bar{g}^{tr}\partial_t\bar{g}_{rr}\notag\\
\bar{\Gamma}^t_{tt}&=-\bar{\Gamma}^r_{rt}
=\frac12(-1+\phi_1)\partial_t\phi_1+\frac12\partial_r\phi_1\notag\\
\bar{\Gamma}^{\theta_i}_{\theta_i r}&=\frac{1}{r}.
\label{christoffels}
\end{align}
The current in eq.~(\ref{Maxwellapp}) is then found to be
\begin{equation}
j^\mu=(\nabla^2_{\rm M}\phi_2,0,0,\ldots,0),
\label{jmuapp}
\end{equation}
where
\begin{equation}
\nabla^2_{\rm M} f(r)=\frac{1}{r^{d-2}}\partial_r\left(r^{d-2}\partial_r f(r)\right)
\label{nabla2r}
\end{equation}
is the Minkowski-space Laplacian of a spherically symmetric function
in $(d-1)$ spatial dimensions.

\bibliography{refs.bib}
\end{document}